\documentclass[journal]{IEEEtran}
\usepackage{graphicx}
\usepackage{nomencl}
\usepackage{url}
\usepackage{multirow}
\usepackage{changes}

\usepackage{hyperref}
\usepackage{float}

\begin{document}

\title{Recent trends in Social Engineering Scams and Case study of Gift Card Scam }

\author{Rajasekhar~Chaganti, Bharat~Bhushan, Anand~Nayyar, Azrour~Mourade


\thanks{Rajasekhar~Chaganti was with University of Texas at San Antonio, San Antonio, Texas 78249, USA. e-mail: (Raj.chaganti2@gmail.com)}
\thanks{Bharat Bhushan was with Department of Computer Science and Engineering,
School of Engineering and Technology (SET), Sharda University, India (bharat\_bhushan1989@yahoo.com)}

\thanks{Anand Nayyar was with Graduate School, Duy Tan University, Da Nang 550000, Vietnam
Faculty of Information Technology, Duy Tan University,, Da Nang 550000, Vietnam
(anandnayyar@duytan.edu.vn)}

\thanks{Azrour Mourade was with Computer Sciences Department, Faculty of Sciences and Technics, Moulay Ismail University, Morocco (azrour.mourade@gmail.com)}


}


\maketitle

\begin{abstract}
Social engineering scams (SES) has been existed since the adoption of the telecommunications by humankind. An earlier version of the scams include leveraging premium phone service to charge the consumers and service providers but not limited to. There are variety of techniques being considered to scam the people due to the advancements in digital data access capabilities and Internet technology. A lot of research has been done to identify the scammer methodologies and characteristics of the scams. However, the scammers finding new ways to lure the consumers and stealing their financial assets. An example would be a recent circumstance of Covid-19 unemployment, which was used as a weapon to scam the US citizens. These scams will not be stopping here, and will keep appearing with new social engineering strategies in the near future. So, to better prepare these kind of scams in ever-changing world, we describe the recent trends of various social engineering scams targeting the innocent people all over the world, who oversight the consequences of scams, and also give detailed description of recent social engineering scams including Covid scams. The social engineering scan threat model architecture is also proposed to map various scams. In addition, we discuss the case study of real-time gift card scam targeting various enterprise organization customers to steal their money and put the organization reputation in stake. We also provide recommendations to internet users for not falling a victim of social engineering scams. In the end, we provide insights on how to prepare/respond to the social engineering scams by following the security incident detection and response life cycle in enterprises.

\end{abstract}

\begin{IEEEkeywords}
Social Engineering Attacks; Gift card Scam; Fake Toll-free numbers; Internet Infrastructure; COVID-19 pandemic; IRS Scam; Phone number spoofing 
\end{IEEEkeywords}

\IEEEpeerreviewmaketitle

\section{Introduction}
\label{sec:intro}
The communication technologies advancement and progression over the years has enabled human beings from all over the world even living in rural areas to get connected and communicate one and other. The communication media can be text, image, audio and video form and the different communication channels like telephone network, computer networks can be used to exchange the information. The user identities should be verified/identified in some manner to enable the communication between two end users using the available technology. The technological user identities may be phone number, email address or social network profile address are used to recognize the users. Although the technology usage for user remote communication improves the quality of the life and remove the distance barriers between people, these technologies pose new threats such as identity theft, social engineering attack/scams, data breaches in organization, ransomware and malware infection etc. \cite{Bidgoli2017} \cite{Tahirkheli2021}. Some non-profit organizations, security service providers and government entities help to protect the user devices and improve the security and privacy of the end users. These security protection services or products at least implemented and followed by organizations so that to align with compliance, protect employees from malicious attempts and maintain their businesses. However, the normal user hardly think about security on a daily basis when using the technology devices. So, they are most likely be the victims of the malicious attempts including social engineering attacks (SEA) \cite{Heartfield2015}. The normal user exploitation is even easier if the user is not at all aware of how the technology works.

Social engineering attacks are well known to be used to target individual users, as the individual users are influenced to become a victim of the social engineering attempts \cite{Salahdine2019}. An adversary may send the phishing emails or deploy maladvertisements in daily browsing websites to perform the social engineering attacks. These attempts may be stopped by security tools in user devices if they have installed security detection tools in the devices. So, the adversary may rely less on technology dependent SES to achieve the malicious tasks. The main difference between the SEA and SES is that SES are mainly performed to manipulate the users with human persuasion rather than weaknesses in technology, even though technology can be leveraged during the execution of SES. As the SEA problem is well explored in literature and various technology solutions based on the machine learning, deep learning, end point security are proposed for addressing the issue \cite{Lansley2020} \cite{Lansley2019}, we focused on SES and herein the social engineering exploitation based scams are discussed in the paper. These scams are very well organized to target section of people. The scammers usually establish the scam life cycle arrangements such as the user manipulation stories preparation, let the users send money to their mule accounts or buy gift cards to share with them, and arrangements to transfer the money internationally in scammer accounts prior to performing scams.

\begin{figure}[!h]	
\centering
\includegraphics[width=8 cm,height=7cm]{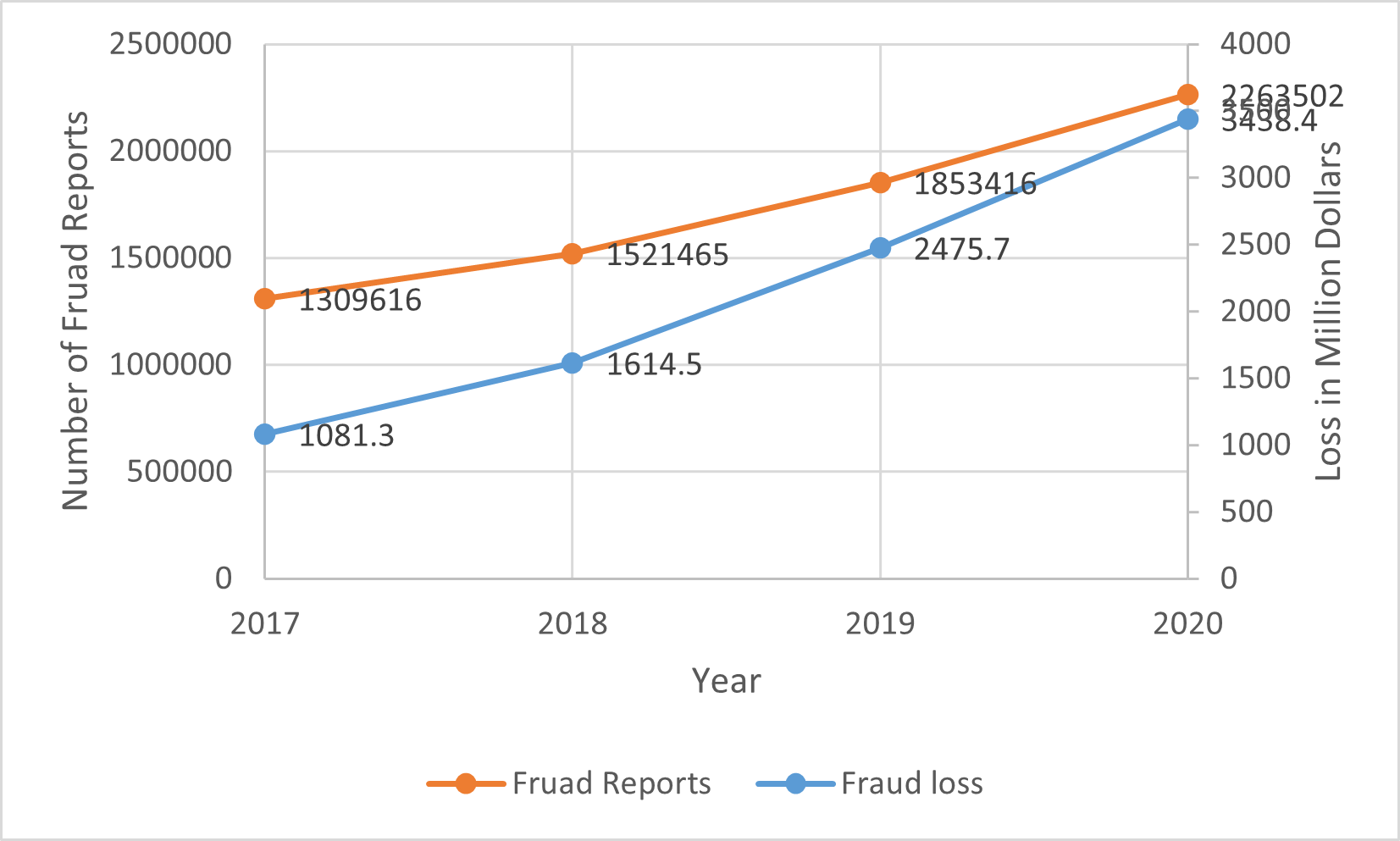}
\caption{Total fraud reports received by FTC in US} 
\label{f.fraudloss}
\end{figure}

Based on the US scam statistics in recent years, the number of reported scams has been increasing year by year \cite{FTC2020}. The scammers always trying to find new avenues to scam the people and most likely get succeed if the victim is not aware of it. As per the Federal Trade Committee (FTC), the total number of the frauds reported by the US citizens increased from 130966 to 2263502 during the four-year period 2017 to 2020 \cite{FTC2020}. We can clearly see in Figure \ref{f.fraudloss} that the fraud reports increases steadily year by year. Furthermore, not to forget, most likely the social engineering scam victims may not report to the government organizations due to the lack of awareness. So, the total number of scams can probably even more than what officially reported in the government public website. Similar to fraud report trends, the fraud loss has estimated to be increased from 1081.3 to 3438.5 million dollars during the four-year period 2017 to 2020 \cite{FTC2020}. The mammoth fraud loss statistics shows that the social engineering scams need to be taken seriously and prevent these scams happening again in the future. 

\begin{figure}[!h]	
\centering
\includegraphics[width=8 cm,height=7cm]{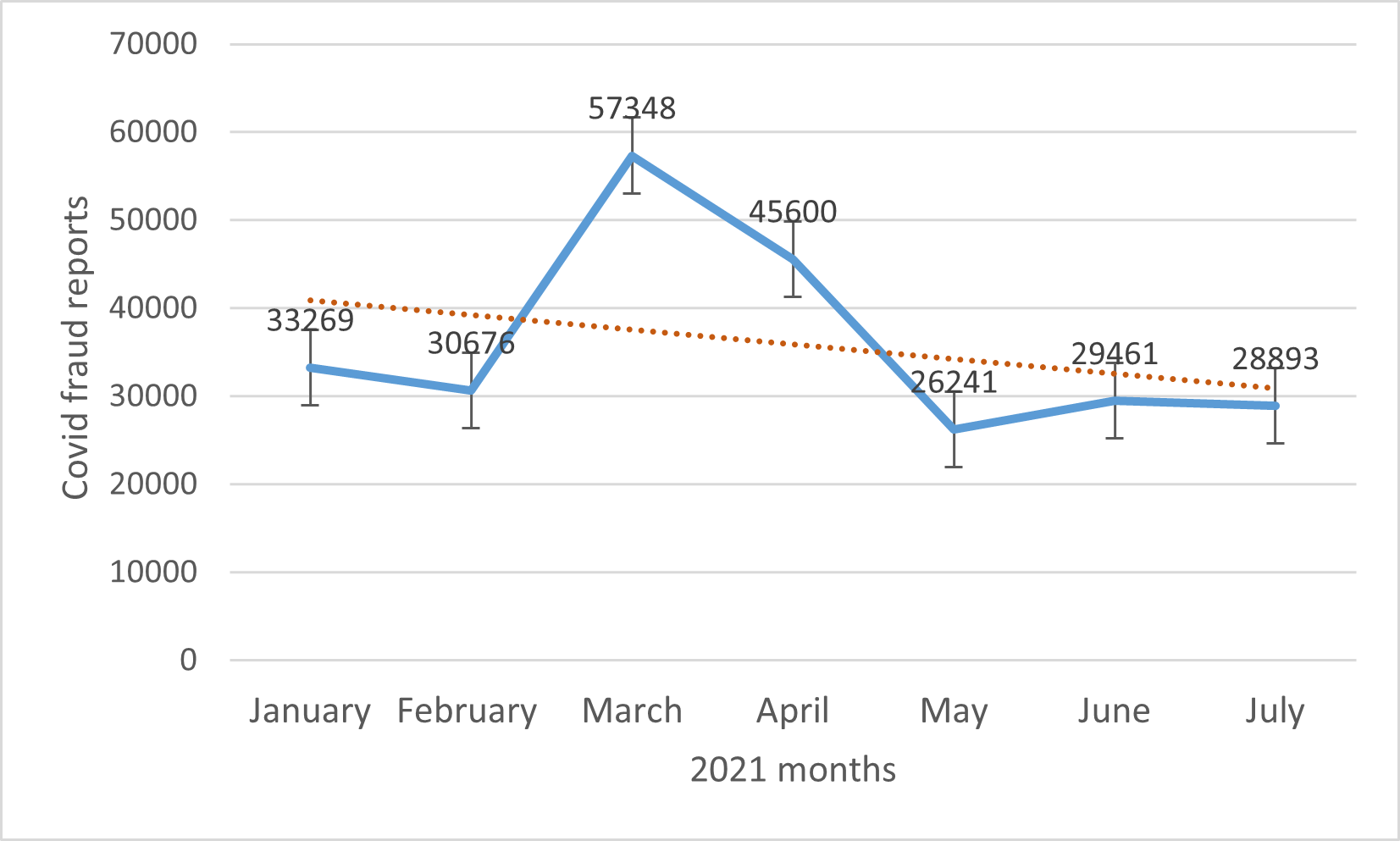}
\caption{Total Covid-19 theme based monthly fraud reports in US} 
\label{f.fraudreports}
\end{figure}

The SES has been evolving constantly and scammers particularly look for current trend social causes or events influencing the normal users to carry out the scams. Notably, the COVID-19 pandemic has been a sensation all over the world in 2020, impacted many lives and changed the human lifestyle.  In particular, the work from home scenario for employees has been a norm in almost every organization during the pandemic. Scammers may take advantage of the COVID-19 pandemic circumstances to target innocent people. To support the claim, we can see that the number of covid fraud cases reports by US citizens is more than 25000 in every month of the first half of 2021 \cite{FTC2020}. As shown in the Figure \ref{f.fraudreports}, the maximum number of fraud cases reported in the month of March is 57348. Furthermore, the reported covid fraud loss is estimated to be \$545 million US dollars so far in 2021. So, these COVID-19 pandemic fraud statistics show that the social engineering scams keep evolving and may see the new avenues for scams in the future. 

As discussed in previous paragraphs, most of the existing research works focus on social engineering attacks discusses technical phishing detection and mitigation solutions, and the impact of the cyberattacks in enterprise organizations when social engineering attack strategies are used for compromising the employee machines \cite{Butnaru2021} \cite{Cui2017}. Even though these solutions may help to prevent phishing attacks or enterprise attacks to some extent, the main social engineering exploitation techniques focusing on the human persuasion is still needed to addressed. There are only few works discussing the human aspects and behavior to understand how the social engineering scams target the individuals. 
The few existing works on the scams mainly focussing on investigating any particular scam, and determining the impact, behavior, characteristics and other essential information about the scams \cite{Bidgoli2017} \cite{Clark2013} \cite{Kharraz2018}. These works solely concentrate any particular scam and may not be used to understand the scam patterns and correlation between different scams. 

In addition, based on our knowledge, the existing literature also mainly focused on addressing the scammer initiated phishing or social engineering attacks and scams \cite{Rauti2017} \cite{Miramirkhani2016}. These attacks include phishing emails, vishing, smishing initiated by the scammer. But, very few works in the prior art explored and discussed on the victim initiated connection to the scammer, and then being fallen as a victim for the scams \cite{Park2014} \cite{Yen2016}. For example, victim visits a social network site and pick a scammer phone number from maladvertisements on the website. Then, the victim may call the scammer and become a target to the scammer. So, overall, in order to give an overview of the human persuasion based social engineering scams rather than technical aspect of the phishing attacks, and also considering the victim initiated scams, we perform a comprehensive  review of the  recent social engineering scams including the recent covid scams and victim initiated attacks. Furthermore, the real time gift card scam targeting organizations attack life cycle is described with examples to provide security awareness among the internet users and consumer users. The tools used in the investigation are also described to enable even a normal internet user can identify these scams with minimal technology skills. We also described the security incident and response guidelines for enterprise security teams to prepare for mitigation of gift card scams targeting enterprise customers and indirectly enterprise organization with internet presence. To the end, the main contributions of work is as follows.

\begin{itemize}
  
  \item This paper discusses the recent social engineering scam trends covering various scam types and the detail description of their operations.
  \item This paper also describe the latest scam trends utilizing the COVID pandemic for performing social engineering scams.
  \item This paper proposes a social engineering scam threat model architecture to map any scam with network components so that  one can understand the scam life cycle for crime investigation and monitoring.
  \item This paper presents a real time gift card scam case study targeted on organizations and provide the guidelines for security incident and response teams to handle the scams.
  \item  This paper also describe the scam threat model for recent COVID unemployment fraud scam and classic technical support scam as a classic examples.

\end{itemize}

The remaining sections of this paper are organized as follows: Section \ref{sec:Literature} includes the literature survey of social engineering scams and attack/scam surveys. Section \ref{sec:threatmodel} includes the threat model for social engineering attacks/scams life cycle with examples. Section \ref{sec:recenttrends} shows the detail review of various social engineering scams. Section \ref{sec:giftcard} and \ref{sec:incident} presents the case study on gift card scam and the guidelines for security incident and response.  Section \ref{sec:conclusion} concludes the paper.

\section{Literature Review}
\label{sec:Literature}
Social engineering exploitation still has been a part of the cyber crimes, even though the technology has been advanced and defenders using advanced technologies like machine learning, deep learning to detect the fraud detection and prevent the cyber crimes. Even in organization security, humans are the weakest link to compromise and perform cyberattacks against an organization. In social engineering, the human behavior/emotions, perception towards the things and tendencies are exploited to successfully perform the attack's and breach the organization. The same social engineering human persuasion is performed to scam the individual users as well.

Firstly, we review the previous works on different social engineering scams and then we compare our contribution in the paper with existing relevant social engineering attack and scams review works. George et al. \cite{Brandon2009} revisited the famous Nigerian check scam in 2009. The authors discussed slight variation of the classical check scam, in which the scammer offers to buy a product and send a check intentionally with payment more than the price of the product. Then, the scammer may convince the victim that more than the price of the product sent in the check by mistake. The scammer also convince the seller to wire transfer the difference pay amount. Later, the check will be determined as bogus and the victim has to pay the check bill when the victim visit the bank for check cash out. The authors also provided recommendations for victim and banker to quickly react to those scenarios. Youngsam et al. \cite{Park2014} set up a honeypot magnetic advertisements in Craigslist to analyze Nigerian scammer behavior, location and other details when the scammers give the response to the posts in email. The scammers send emails to the assumed victims (researchers) in response to their posts and then the emails header and content information is analyzed to identify the behavior of the scammers. These advertisements are set up particularly targeting scammer automated scripts running when responding to the advertisements and intentionally keeping high prices in Craigslist to let the legitimate users not opting to buy the products. The Nigerian scammer groups and their activity signatures are identified in this work using honeypot advertisements.  However, this work is limited to scammer responding to the victim/user advertisements in Craigslist. Marzuoli et al. \cite{Marzuoli2015} created a honeypot to collect the robocalls and spam calls originating from the scammers. Then machine learning techniques applied on the robocalls first few seconds of the audio to cluster the scammer groups. The bad actor fingerprints also determined to distinguish the scammer group. However, the study on small scale robocalls may not cover all the telephone based scams and groups behind them. Ting-Fang et al. \cite{Yen2016} discussed various romance scams, including luring the dating app victim to visit third party sites by clicking advertisements, sending romance emails and  performing romance phone calls. By using the simulated spam filter, the authors able to determine that the scammers respond to 2\% of the received auto generated emails. 

Najmeh et al. \cite{Miramirkhani2016}  performed a detailed study on the technical support scams by automatically crawling the maladvetisements on the internet and collecting the scam phone numbers and URLs for further study. Furthermore, validation of the URL and phone numbers is performed for detecting the technical support maladvertisements. Later on, 60 different scammers are selected to talk with them and collected all the details including  scam process and tactics used by scammers for performing the scam. Park et al. \cite{Park2017} performed an empirical assessment of the Craigslist rental scams. The Craigslist rental scam posts are analyzed using automated bot and then classified posts into group campaigns. Additionally, the authors interacted with the scammers to identify the infrastructure needed for operating this scam. Overall, the authors deduce that considering credit card payment system protection may reduce these scams in the future, as they mentioned that 80 percent of the rental scams relying on the credit card transactions to steal the money from the users. Vidros et al. \cite{Vidros2017} introduced automated recruitment fraud detection using natural language and machine learning techniques. The recruitment dataset of 17,880 annotated job ads are collected and released for public. Srishti et al. \cite{Gupta2018} performed a study on large scale social media campaigns used to distribute the scammer phone numbers in the social media networks. They identified that 202 campaign groups actively posting phone numbers on the web and also mentioned that some social network (twitter) can red flag as spam campaign better than the other popular social  media network (Facebook). The authors recommended the necessity  of sharing the threat intelligence data among the social media networks for early detection of spams.

Hu et al. \cite{Hu2019} presented a systematic study on the fraudulent dating apps. Millions of apps collected from the android play store and then performed static analysis to identify the in-app purchase apps. The identified apps are then clustered into group of app families. These app families are manually inspected and also analyzed the user comments on the apps to determine the fraudulent dating apps. These fraud apps use the chatbots to influence the users buying the premium services. Suarez et al. \cite{Suarez-Tangil2019} proposed automatic detection of online dating fraud user profiles. The scraped user profile are used to extract the demographic, image and profile description content. Then, the features are created for each of these categories. The ensemble based machine learning technique is used to classify the given profile is legitimate or fraud using those features. The authors reported that the optimal results obtained for the experiments performed on the user profile features using ensemble methods. Agari \cite{Agari2019} solely described on the phishing romance scams and mentioned that the targeted category of the people are becoming the victims of this scam. They also mentioned that divorced, farmer and disabled people are mostly being impacted by these attacks.

Pastrana et al. \cite{Pastrana2019} presented a semi-automatic analysis on ewhoring scams using the underground forums. The machine learning and natural language processing solutions are proposed to extract the threads posted on underground forums. Then, the extracted URL and images from the threads are considered to find out child abuse material. The reverse image search analysis, domain lookup tools are used for finding ewhoring cases. 

The review of the above prior art scams reveal that few of the scams such as dating scams, romance scams, technical support, Craigslist rental, online dating apps scams are studied and performed in depth analysis for identification. However, some of the scams such as law firm scam, Cat scam, Gift card scam, Mortgage scams detection and analysis are not covered in the state of the art. So, we wanted to address this issue by discussing all the recent social engineering scams, and provide awareness and recommendations to handle these scams. Furthermore, our work can be used as a reference to perform the research on the research gaps in recent social engineering scams.



\begin{table*}[]
\centering
\caption{social engineering scams categorization } \label{T:scamtypes}
\resizebox{\textwidth}{!}{%
\begin{tabular}{|l|l|l|l|l|}
\hline
\textbf{Author} & \textbf{Scam Name} & \textbf{Technique} & \textbf{Advantages} & \textbf{Comments}     \\ \hline
George et al. \cite{Brandon2009} & Nigerian check  & larger amount check & The check scam protection discussion  &  Only two ways of performing check scam mentioned \\ \hline

Youngsam et al. \cite{Park2014} & Nigerian Craigslist  & magnetic honeypot advertisements & better understand Nigerian scammer patterns, tools, email usage etc.  & 10 groups responsible for most of the activity \\ \hline

Aude et al. \cite{Marzuoli2015} & Telephone scams & Honeypot, Audio machine learning & Able to group the scammers & Identified one third of the calls are robocalls \\ \hline

Ting-Fang et al. \cite{Yen2016} & Romance scams & Simulated spam filter to track the scammers & Discussed different types of romance scam & Affiliated market scams using online dating apps \\ \hline

Najmeh et al. \cite{Miramirkhani2016}  & Technical support & Discovering advertisement in web & Automated discovery & Still considered as most dangerous scam \\ \hline

Youngsam et al.\cite{Park2017}  & Craigslist Rental & Web crawling, automated responses  & Able to identify the scammer infrastructure used  & less than half of the scams Craigslist identified and removed from portal \\ \hline

Vidros et al.  \cite{Vidros2017} & Recruitment fraud & Automated web page crawling  & first dataset available in public  & ML based techniques used for detection \\ \hline

Srishti et al. \cite{Gupta2018} & Spam campaigns & Automated crawling of webpages & large scale study covering posts on multiple social networks & Twitter can suspend spam accounts better than Facebook \\ \hline

Yangyu et al. \cite{Hu2019} & Dating App scam & 
Crawling dating apps in Android store & Large scale analysis and fraud app detection & The chatbot accounts influence the users buy premium \\ \hline

Suarez et al. \cite{Suarez-Tangil2019} & Dating fraud & ML classifier & Achieved 96\% correct identification &  Online profile based analysis efficient compared to bot based \\ \hline

Agari \cite{Agari2019} & Romance scams & Phishing emails & - & target divorced, farmer, disabled people\\ \hline

Pastrana et al. \cite{Pastrana2019} & eWhoring scam & Crawl Underground forum & 
Pipeline framework to identify these scams & Performed URL and image analysis during the investigation \\ \hline

\end{tabular}%
}
\end{table*}

In order to show the uniqueness of our work and contributions, we compare our work with the social engineering related works in the prior art. Chitrey et al. \cite{Chitrey2012} conducted a social engineering attack survey to understand the IT service provider employees and students from top IT colleges perception on the attacks. The authors mentioned that the obtained results can be used to develop the information security policy and security awareness programs for organizations. Tu et al. \cite{Tu2016} provided a detail review of the existing telephone scams and evaluated the existing telephone scam solutions in the state of the art. They concluded that there is no universal solution to stop the telephone scams. So, based on the desired requirement like usability, deployability, and robustness, the solutions can be implemented and combat these scams. In contrast to our work on scams, tu et al. \cite{Tu2016} mainly focused on the telephone scams.

Anjum et al. \cite{Shaikh2017} performed a review on the phishing attacks and anti phishing schemes for mitigating the phishing email. However, the review is limited to phishing emails and none of the scams are discussed in the article. Fatima et al. \cite{Salahdine2019} performed social engineering attacks survey. They described various social engineering attacks, prevention and mitigation techniques. Additionally, various computer based countermeasures and mitigation techniques are compared to identify the advantages and limitations of each technique. However, the paper only covers attack originating from the scammers and social engineer attacks needed technology to execute the attacks. 

Yasin2019 \cite{Yasin2019} performed literature review of social engineering attack and human persuasion methods used for those attacks. The authors described that thematic and game-based analysis techniques are effective to better understand the attack scenarios. The empirical evaluation gives neutral results on the game based analysis of security awareness assessment. 

Alzahrani et al. \cite{Alzahrani2020} discussed the coronavirus social engineering attacks and recommended security awareness is the solution to mitigate these attacks. Although the author's work can be used as a good reference for covid social engineering attacks, the article did not describe the detail description of the covid social engineering scenarios.

The authors \cite{Alabdan2020} performed a detailed review of phishing attacks, which covers the different attack types, attack vectors and communication medium. They also described the anti phishing methodologies to mitigate the attacks and future challenges. However, the paper did not address the victim initiated scams used to let the victim reach out to the scammer and getting scammed. Venkatesha et al. \cite{Venkatesha2021} emphasized the social engineering attacks leveraging the COVID pandemic. The authors discussed the social attack trend shifting as the COVID pandemic emerged, and discussed the COVID thematic attacks. However, all the covid based attacks/scams are not covered in the article. In \cite{Akdemir2021}, the authors performed an analysis on the COVID theme phishing emails. These COVID based phishing emails used authority, liking and commitment as the principles in those emails. The intention of these emails is to let the victim respond back to the scammers so that they could demand the money as part of the scam. Additionally, the personal information also being gathered in the email based phishing scam.

\begin{table*}[!h]
\centering
\caption{State-of-the-art comparison with our work} \label{T:priorreview}
\resizebox{\textwidth}{!}{%
\begin{tabular}{|l|l|l|l|l|l|l|}
\hline
\textbf{Author} & \textbf{Scam/Attacks}    & \textbf{Article Focus} & \textbf{Advantages} & \textbf{Comments}         & \textbf{Case study} & \textbf{Covid Scam}           \\ \hline
Chitrey  et al. \cite{Chitrey2012} & Attacks & Employee and student survey & Collecting the user perception samples on SE  & limited to India & - & -\\ \hline

Tu et al. \cite{Tu2016} & Scam & Review of telephone scams & Detailed overview of all telephone scams & No universal solution for telephone scams & - & - \\ \hline

Anjum et al. \cite{Shaikh2017} &  Attacks & Phishing email based survey & - & limited to phishing emails & - & -\\ \hline

Fatima et al. \cite{Salahdine2019}  & Attacks & Social engineering attacks survey & Review of the most technical SE attacks & The human aspects of SE is not covered &   -  & - \\ \hline

Yasin2019 \cite{Yasin2019} & Attacks & SE attacks and human persuasion methods review & Many attacks in literature are covered in the review  & game-based analysis on SE produced neutral results & -  & - \\ \hline

Alzahrani et al. \cite{Alzahrani2020} & Generic & Corona phishing survey & Recent trend attacks and scams discussion & limited to coronavirus SE  & -  & Yes\\ \hline

Alaban et al. \cite{Alabdan2020} & Attacks & Phishing attacks review and countermeasures & Detailed survey on SE attacks/scams & Only covers scammer initiated attacks & - & -\\ \hline 

Venkatesha et al. \cite{Venkatesha2021} & Generic & Covid thematic & Discussion on COVID attacks/scams & Only Covid thematic attacks/scams covered & -  & Yes\\ \hline

Akdemir \cite{Akdemir2021} & Generic & Content analysis of Covid emails &  Analysis of covid scams & Only phishing Covid emails discussed & - & Yes \\ \hline
Our work & Scam & Review of scams and case study & Threat model to map any SE scam/attack & - & Yes & Yes\\ \hline
\end{tabular}%
}
\end{table*}

Overall, based on our review of the proposed solutions for combating the social engineering scams, and the review of the existing social engineering attacks and covid social engineering surveys, there is clearly a gap on discussing the recent attack or scam trends targeting innocent people, and particularly the review of the recent scams is not discussed in the prior art. The recent works rather focused on the technical aspects of the social engineering attacks. We try to address the research gap and aim to present an overview of the scams with emphasis on the human persuasion. Additionally, the threat model for social engineering is not discussed in the prior art. We have presented a threat model architecture for addressing the social engineering attacks.
\section{Threat Model}
\label{sec:threatmodel}

Social engineering scammer identification is much more challenging than one can think of, as the scams involve intricate steps to trace out and multiple stakeholders involves during the scam lifecycle. So, a common language and criteria is required to identify the scammers and reduce the scope of the scammer presence for catching them. Another reason for identifying the scammers are difficult because they use combination of technologies like telecommunications, internet, social media, web searches to execute their plan of actions. It is also challenging to have security workforce, who are proficient in multiple disciplines to track the scammers and work with multiple stakeholders. Another challenging task is the jurisdiction issue when professionals working internationally to catch scammers located in remote countries. In order to understand the terminology and techniques used in social engineering scam life cycle, we have proposed a social engineering threat model architecture, which can be used to map any social engineering scam happening anywhere in the world starting from scammer origin to victim location. The proposed social engineering scam threat model is inspired by telephony and internet architecture \cite{Sahin2017} \cite{Mcinnes2018}.

\begin{figure*}[!h]	
\centering
\includegraphics[width=14 cm,height=8cm]{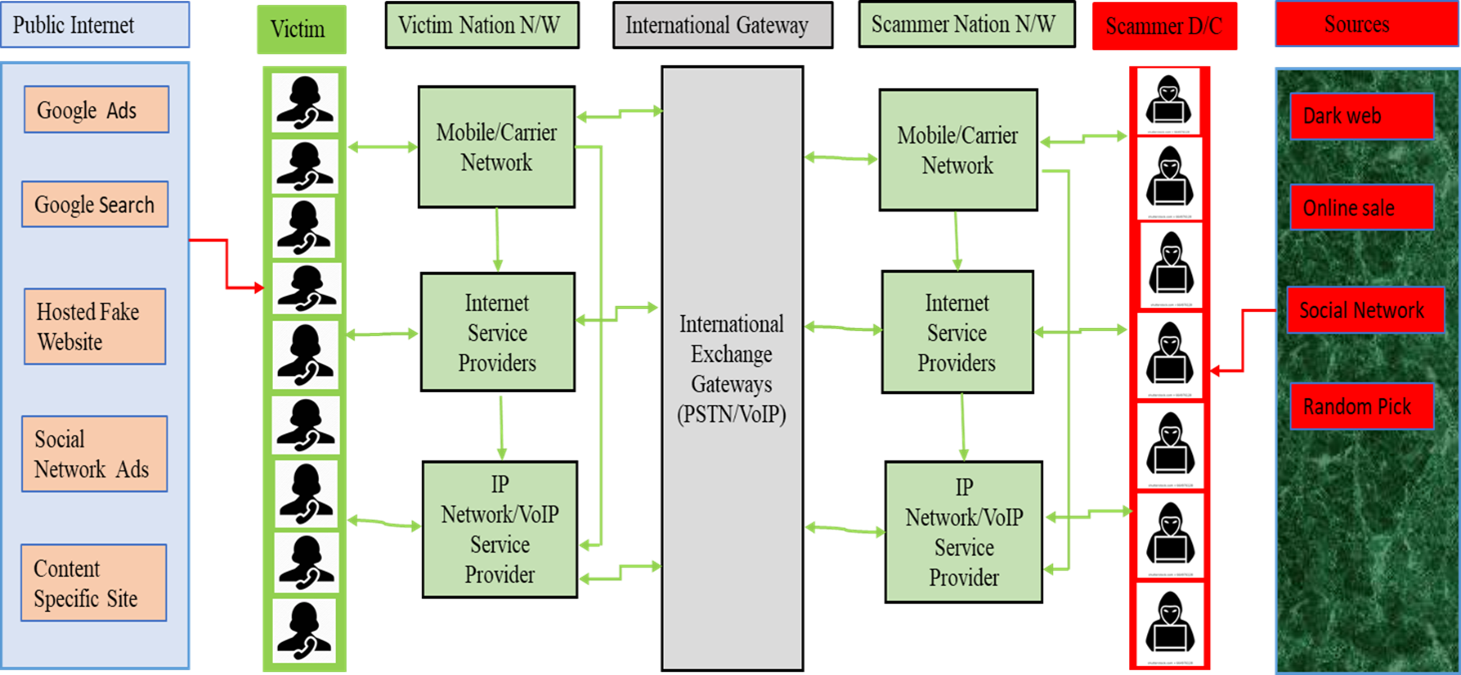}
\caption{Social engineering scam threat model architecture} 
\label{f.threatmodel}
\end{figure*} 

As shown in Figure \ref{f.threatmodel}, the social engineering scams threat model architecture mainly consist of the victim national communication network, international gateway, scammer national communication network and scammer data center modules. Additionally, the data collection sources, public internet resources are mentioned to be used in scam life cycle. The detail description of each component in the module are discussed to understand the role of each component in the social engineering scam life cycle.  \\

\subsection{Public internet:} 
Scammers get abundant information regarding the people to scam from internet. As the social media applications usage has grown significantly over the years, users share lot of private information in these sites, and don’t consider the privacy is an issue. Scammers collect targeted user group information from these sites. Furthermore, some websites provide services to sell the personal information for profit. So, it is not difficult for the scammers to create the user database with personal information like phone number, occupation, name etc. For instance, the Facebook and Instagram social media sites publicly provide the users data, who don’t set their privacy settings to private \cite{Syn2015}. A scammer may run a web crawler with API calls to collect all the targeted user’s information and save it in a database. Some of the essential technologies on the internet are used for social engineering are described below. \\ \\
\textbf{Google Ads:}
Google offers Ad service to customers, who wants to promote their products and services \cite{Lee2011}. A traditional way of showing Google Ads in Google search results based on the keywords used in the search. The Google Ads usually show the service or product website and phone numbers when the relevant keywords used for search. For instance, if a user search for “home rentals near me” in Google search, the number of results will appear along with Google rental ads. Scammer may use the keywords like “home rentals” and “near me” to pop up their fake website or their phone number to attract the users contacting them. So, Google Ads can be used a secret weapon in social engineering scams to let the user calling the scammers and scam the users.  \\ \\
\textbf{Hosted fake website:}
The scammers misdirect the users using fake web applications deployed in public internet. They may post illegitimate information or scammer contact number to contact the scammers. Scammer may simply need domain name and domain registration with registrar to start using the domain for hosting the applications. Based on the targeted victims, they may post the relevant content to attract the victim to visit their pages and then direct them to achieve the intended goal. Most of the time, the fake website is used to provide the scammer contact details in scams so that the user/victim can reach out to them, and then execute the scam. Later, the scammer may ask them to wire transfer the money or buy gift cards from store and then the scammers use them for buying goods etc. So, based on the scammer goal and their strategy, they instruct the victims and direct them to involve them in scams. \\ \\
\textbf{Social network Ads:}
Google Ads may be having limited visibility and few daily visitors. This may be due to few of them would like to focus on ads when searching for any content. But social network Ads has larger reach, as most of the social media users spend time for fun and most likely they will explore the Ads for entertainment purpose. So, social media Ads has got more attention recently and it is convenient to reach out many people \cite{Alalwan2018}. Scammers started leveraging the social media Ads to attract the users and perform the scams. For instance, to scam the people who love cats, the scammer may populate the cat related post Ad pages with cat scam relevant social media Ads. Users are most likely to visit and explore these Ads to buy a cat. The scammers can now scam them to pay money for cat online order and shipping to their location. Later, the scammers may disappear once money is received. Similarly, there are many other types of scams performed by posting fake Ads in the social media, as discussed in section \ref{sec:recenttrends} \\ \\
\textbf{Content Specific website:}
Scammer may also focus on the specific targeted users for financial gain, popularity, or political advantage to their clients by posting the specific content in targeted websites. For instance, during the elections, the scammers may host fake website hosting content supporting a particular presidential candidate and indirectly support them to gain more votes by the influencing the voters who visit the fake websites. The scammer may benefit indirectly, possibly some supporters pay them money for hosting the content and supporting their candidate. Although the people not directly involved to be a victim of the attacks, Scammers still earn money by indirectly influencing them. \\ 
\subsection{Victim:} Users are being targeted by scammer and few of them are really being impacted by the scams. All the impacted users will be the victims in the context of our social engineering scam threat model architecture. The communication can be initiated by the victim to visit the Google Ads or social website Ads, or the scammers may phone call or email the victims \cite{Gupta2015}. \\ 
\subsection{Victim National Communication Network:}
Victim national communication network mainly comprised the victim mobile/carrier network, internet service provider network and VoIP service provider network to connect with the scammers resided in the victim living country or international connection to the scammer country. It is essential to understand that the victim resided communication network components in detail to better protect the users and implement the security mechanisms \cite{Sahin2017}.  \\ \\
\textbf{Mobile/Carrier network:}
Mobile network involves in the social engineering scam when the scammer connected to the user through phone call/messages. For example, the main carriers like AT\&T, T-Mobile, Verizon in the United States only process the messages/phone calls when the victim is using the same set of carriers and the scammer is trying to send a message or call them. The network connection could be LTE, 3G, 4G or 5G to perform telecommunication operations. However, security monitoring at the carrier level may be ineffective for scam call detection, as the legitimate call and scam calls are difficult to distinguish. It is very challenging to implement any security solutions for social engineering scam detection at the mobile network level. So, it is very unlikely that the mobile networks can help to prevent the scams. However, if the scammer calling from spoofed phone number, A recent proposal of STIR/SHAKEN protocol implementation at the carrier network level may authenticate the caller ID and prevent the robocalls. This may reduce to an extent the scammers who rely on caller ID spoofing for social engineering scams. But the protocol implementation may not completely prevent the scams. Mobile networks can also help the users to detect the phishing or scam messages. Currently, the users still receive the phishing or scam messages in the mobile phones, and detection and prevention of the scam message solutions are required. We have not described the technique used for performing the telephone scams in the mobile network targeting operators, third party service provider etc. the detail description of those scams can be found here \cite{Sahin2017}.  \\ \\
\textbf{Internet service provider}
There may be instances where the users may rely on the wireless internet connection to complete the phone calls. These calls are processed as VoIP packets and passed through the user internet service provider network prior to forward to the core network and international gateway. Although there is more visibility on the VoIP packets passing through the internet service provider,  the packet level analysis is unlikely helping to identify if the call is scam call or not. So, security controls implementation for scam mitigation or prevention may not be a viable option at internet service provider level and are extremely unlikely to detect scams at the internet service provider level. \\ \\
\textbf{VoIP service provider} 
Voice over Internet protocol (VoIP) services \cite{Mcinnes2018} are most likely provided by third-party network like Bandwidth.com to connect the international VoIP calls. The scammer may leverage these services to obtain the victim country phone numbers and may use it to call forward the international calls. These services make it difficult for an end user to identify the called person location. The STIR caller identity protocol implementation may alleviate some of these scam problems with identity authentication, even though it won't completely eliminate the scam issues. \\ 
\subsection{International exchange gateways}
The international exchange gateways forward internet or telecommunication connections from one country to other country. For instance, the optical fiber-based network connects through sea for communicating between the two networks. As the traffic rate is huge and enormous, the scam related network traffic monitoring is almost impossible. Furthermore, gateway operator or owner is not involved in scams. So, they have little or no interest to even consider the scams are really a concern. In general, the scam detection implementation is impossible and practically not a viable option, as the scams are mostly based on the human persuasion rather than fixing the technical problem. So, we may not rely on preventing the scams at the international exchange gateway level. \\ 
\subsection{Scammer call center}
Scammers usually operate as a legitimate company managing call center services. Based on the historical scam events, most of the call centers are known to be operated from India or Nigeria. For instance, the Indian police busted a call center cheating Americans for 14 million dollars recently \cite{Dhillon2021}. They reported that the call center hires the employees and train them to talk in American accent. The scammer employees are also instructed to following vetted script while talking to Americans. They basically use various scam scripts to obtain gift card or wire transfer money from victims. Typically, the call centers may contain VoIP gateway setup to perform international calls, computer equipment that a typical call center requires.  \\ 
\subsection{Data Source} 
Scammers collect the targeted user information from disparate sources prior to initiating the scam operations on them. This is the first step and one of the essential steps in scam operation life cycle. If the scammers have more user dataset, they may get more chances to succeed. The user information is typically the phone number, email address to get in touch with the users. The most often used data sources are described here. \\ 
\textbf{Dark web}
Scammers find dark web is the most reliable place for buying the targeted user data illegally. The data should be coming from data breaches, stolen information, unknown disclosure etc. The data can be sold in cheaper prices at times compared to lucrative amount made when the scam is successful. Cryptocurrency may be used to do financial transactions and maintaining the user anonymity. So, the identification of the real person involved in these transactions is almost impossible. Most of the time, the normal user credentials are being sold in dark web after collecting from data breach disclosures. \\ 
\textbf{Online sale} 
A number of third-party services provide user data available for legitimate sale through business websites. The user data may include the phone number, email, username, location etc. However, they may not sell the personal credential, banking information. It is partly easier to get the user data like name, date of birth, location, phone number etc. Additionally, scammers can also freely collect the basic user information in bulk from the legitimate sites, which are maintained for tracking purpose. \\ 
\textbf{Social network}
Social network collects and store lot of user information. So, scammers run slow web crawlers to periodically perform API calls for collecting vast amount of user data for scamming purposes and evading the security detection. The social networks like Facebook and Instagram maintain lot of information and even google search crawler can help to download lot of public user information. For instance, the user details of any particular organization or location can be extracted with simple search queries in google. \\ 
\subsection{Technical support scam threat model}
\begin{figure}[!h]	
\centering
\includegraphics[width=8.9 cm,height=6.5cm]{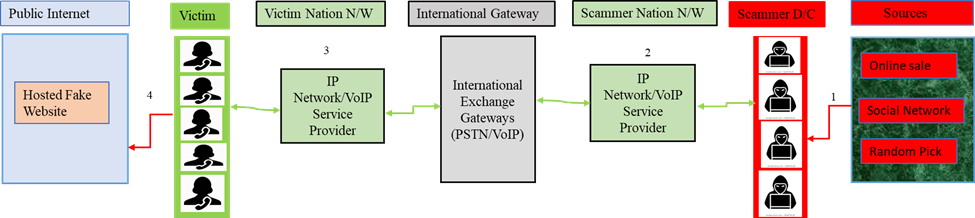}
\caption{Tech support scam threat model} 
\label{f.techsupport}
\end{figure}

The technical support scam threat model defines the various components between scammer and victim while performing scams. In the first phase, the scammers obtain the user phone numbers from multiple sources such as public internet sites, social network profiles, random selection of the phone number using state and area codes. The scammers maintain list of users database and typically recruit the people who are desperate to work for any job. They may maintain a datacenter to perform scamming operations. As we can see in the Figure \ref{f.techsupport}, the second and third phase covers the communication between the scammer and the victims. As the phone calls are being used to reach out to the victims and use VoIP protocol to spoof the phone number, the IP network is being used as a communication network \ref{f.techsupport}. In phase 4, the victim is being instructed to visit the scammer-controlled web application for making them believe that the victim is indeed have a performance or virus issue. The scammer make them believe that their machine is compromised with virus and demand service charges. The victim may wire transfer the money or sending gift cards to scammers to complete the scam based on the scenario. \\ 
\subsection{Covid unemployment scam threat model}
\begin{figure}[!h]	
\centering
\includegraphics[width=8.5 cm,height=7cm]{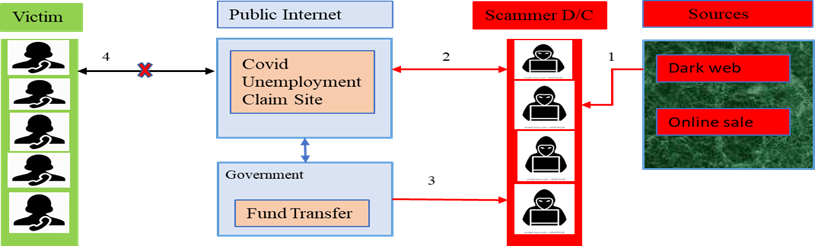}
\caption{Covid unemployment threat model} 
\label{f.covidunemployment}
\end{figure}

The threat model for covid unemployment has a unique characteristic that the victim is not directed contacted by the scammer. So, the communication modules between victim and scammers have been eliminated. As shown in Figure \ref{f.covidunemployment}, the threat model comprise three phases to complete the scam.  Firstly, the US citizen's personal information including the phone number and the social security number is collected from either posted data breaches in dark web or third-party private sales. In this scam, the username and social security number was enough to successfully divert the government unemployment claim refunds to scammer bank accounts. The scammers fill the applications in government unemployment claim web application portal, as shown in the phase two of the model \ref{f.covidunemployment}. The scammer leverages the weaknesses in the web application for identity verification. They may provide the bank account of mules or scammer operated account to collect the money. The unemployment claim funds moved to the scammer bank accounts in phase three. Later, when victim try to fill the unemployment application for COVID unemployment, the web application rejects the case as the claim is already processed by the scammers. \\ 
\section{Recent Social Engineering Scam Trends}
\label{sec:recenttrends}
In this section, various SE scams are discussed in detail including the COVID scams. In addition, the SE scams are compared and classified based on the SE principles, scam initiations, entity involved, locations, technologies involved and targeting victim ages. \\
\textbf{Covid Unemployment Scam:}  
Social scammers quickly adapt their techniques to perform scams on targeted innocent people. COVID-19 pandemic has been a topic for the last one and half years, and people from all over the world are concerned about the wide spread of the virus. Number of people have been impacted with unemployment due to COVID lockdown. The governments standstill to support their citizens and helped to financially support them. Scammers use the unemployment reporting loopholes to fraudulently claim the citizen unemployment funds, and transfer the money to mule’s accounts so that the funds moved out of the country to remote scammer bank accounts. US covid unemployment scam is an example of this category  \cite{Team2020}. Firstly, the scammer collects the US citizen's personal information including the social security number from data breach disclosure sales in dark web or compromising victim machines to steal the personal information or collecting from third-party sellers. Furthermore, additional personal details like email, phone numbers collect from the public websites through web searches. Then, scammers fill the unemployment claim forms through state level government websites. As some of the websites did not actively verify the identity and lack of security measures like multifactor authentication installed on the applications, the scammers use the loophole to claim the citizen's unemployment benefits. The government then release the unemployment funds to assume that the actual citizen claiming the unemployment funds. But, the financial transactions received by the scammers operated mule accounts. Later, the money is wire transferred to international bank accounts without getting caught by bankers. If the US citizen really job impacted with Covid later tries to submit the unemployment claims, the websites report them that the submission already received, and says that the funds are transferred. The innocent citizens are being scammed and has to file complaints to government for support. This is a classic social engineering scam leveraging the latest trends without even the victim involved in the scam life cycle \cite{Farley2020}. As the reports seen from multiple members, the US government was able to quickly identify the scam and fix the website application issues in many states with more protection and identity validation for reimbursement. There have been reports that the stolen money bank accounts have been seized and recovered some amount of money from the scammers  \cite{Farley2020}. Unfortunately, the victim is not involved and most of them not aware of these scams to take precautionary actions.  \\ 
\textbf{Covid Vaccine Scam:} The scammers adapting the scamming strategy according to the covid vaccination scenarios and vaccine development stages since the covid pandemic existed. During the early stages of Covid, the vaccine is not available and still scammers run fake websites offering covid vaccines. Additionally, the robocalls used to reach out the people and lure them with vaccine for money \cite{Alzahrani2020}. Once the money is received, the scammer will disappear from the user radar. The scammers also collect the personal information for data gathering purpose using vaccines scams. Later on, when the covid vaccine is approved, the scammers using the covid vaccination appointment booking fake websites for collecting the information. Overall, as the vaccine transition from developing to available state, scammers changed their strategies to execute scams with content appropriate to the situation \\ 
\textbf{COVID-19 survey scams:} The survey are mainly used to collect the personal information such as names, location, phone number, email etc. The collected information may be used to execute various scams or phishing attacks or selling the data in underworld communities. The scammer create fake covid survey content websites, similar to fake information spreading websites seen during the US election in 2020 \cite{Chaganti2020}. So, the users should be aware of the web links they browse when using the internet. The survey requests may also come through the phone messages. It is recommended not to click the web links received through phone messages so that the used are not being scammed. \\ 
\textbf{Coivd-19 stimulus scam:} 
The stimulus packages offered by US government is also a topic of interest for the scammers. They would like to perform scams using stimulus as a topic and gather personal information. They may ask to request bank account credentials to steal the money. A proper stimulus receiving confirmation is required and only need to talk with the government official when someone approached through phone or emails so that not become a victim of these scams. Scammers may also offer tests, health and human services grants, and medicare prescription cards in exchange for personal details, including medicare information \cite{Venkatesha2021}. \\ 
\textbf{Covid fake donation/charity scam:} The scammers may use the fake donation organization or charity names to collect the money from innocent people. The scammers leverage the goodness of the people and lure them to pay the money for donation. Instead of sending the money to fake organizations, the money is transferred to scammer bank account and then scammers will cut the contact from the victims \cite{Khan2020}. These are also mostly happen through internet or phone calls. So, the security awareness of scams is needed to combat these scams.  \\ 
\textbf{Gift Card Scam}
In Gift card scams, the scammer contacts the victim through social media networks, phone messaging to initiate conversation and make them believe that they are legit. As the victim believes them as legit people, the scammer offers to help them financially in return for buying the gift cards from stores like Walmart, target, CVS etc. If the victim is greedy and not aware of repercussions of the scams, they may end up buying the gift cards from the stores and share the gift card numbers with the scammers. Whenever the scammers receive the gift card number, they disconnect from the victim and never be in contact with victims. So, the victim loses all the gift card money through these gift card scams. The scammer may use different ways to monetize the stolen gift card numbers. If the scammer is operating from another country, he/she may use mules to spend the gift cards for buying things and later the mules pay the money to the scammers. Some scammers may use those gift cards to buy things for themselves and leave no traces to track them. It is highly difficult to track and identify the scammers even if the victim reported to the authorities or gift card bought stores within the short span of time, as tracking the gift card used store, identifying the scammer involves working with multiple stakeholders like legal, ethical, technical teams from multiple entities like laws enforcement, gift card provider etc. Unless it is an organized multi-million dollar scam impacting the reputation of the organizations and lot of people were reported to be a victim of the scam.   \\ 
\textbf{Romance Scam} 
Scammer leverage the people liking tendency towards someone as their weakness for performing scams. The romance scam mainly involves social media networks, dating sites to connect with the victims. Scammers create fake dating profiles with fraudulent information trying to attract people in different age groups. They build a relationship and constantly pursue to impress them for gaining trust among each other. The scammers may even pursue to let them build the sympathy towards them. For instance, the scammer may tell stories involving family issues, jobs issues and  employment issues in their country etc. to get attention and sympathy from the victims. Once the trust is built, the scammer may exploit the victim to borrow money for a favor, financial assistance to help other people etc. to collect the money. When the scammer gained the intended financial help, he/she may disconnect from the partner/victim and delete the social media accounts/phone numbers for disappearing from the victim world. The number of romance cases has been growing recently, as the number of websites/apps increased tremendously \cite{Official2020}. The romance scams have been changing as the technology advances with instance video chat, international dating apps etc. Another type of romance scam is faking the profile as a woman and pursing the men with love and lust to steal the money from them. The scammers build a relationship claiming to be women and then blackmail them to post their content in public. If the victim has reputation in the society, he may pay the money to scammers to save their reputation and get away from it.    \\ 
\textbf{IRS Scam} 
Internal Revenue System (IRS) scam is known to be one of the oldest and well-known scams in the scam history \cite{MansiChoksi2017}. A scammer collects the list of the phone numbers and their personal information from the third-party vendors or public search websites. The aged people contact details are collected for ease of scamming. Once the numbers list obtained, scammer located in other countries like Nigeria call them claiming to be IRS officials. The victim can see the call is coming from the United States, as the scammers use international VoIP phone services to hide the originating phone number. The scammer tries to convince them that they are from IRS and instruct them to pay the fabricated IRS taxes, which the victim is not liable to IRS in reality. If the victim is frightened  or not aware of these scams, he/she may transfer the money to the scammer bank accounts with fear. Once the funds are transferred to the scammer, he/she stop talking to the victim and may elect another victim to perform the fraud. Unfortunately, the victim reaches out to the legit IRS for resolving the issues and the IRS could not be able to help them to retrieve the funds from scammers in this case. The scammers are using new technologies to reach out to the victims nowadays, even though the scam format has not been changed over the years \cite{Bidgoli2017}. So, simply knowing the fact that IRS will not call you or reach out in social media to pay the taxes would prevent the citizens for not being a victim of these scams.   \\ 
\textbf{Check Scam} 
Scammers may use the loophole in  banking system, i.e,  usually take few days to find the fraudulent checks once submitted to perform the check scams. Scammers send you a check with an amount more than you are supposed to be receiving from an entity, and tell you to transfer the overpayment or ask you to pay the money in alternative forms. Initially, the scammers request the victims to draw the cash in check and give fake checks in exchange for money by saying urgent need of money  \cite{Green2020}. The check scam trends have been changing with the advancement of the internet. Now, the scammers may give you legitimate checks owned by mules or legitimate users. Then, they ask the victims to transfer the money and cash out the check in the bank in exchange. Scammers first receive the money through wire transfer, and then when the victim visit the bank, he will be notified that the check will not work. Now, the victim can’t retrieve the money sent through wire. So, it is highly recommended to not accept the checks which value more than the intended price and don’t use the money offered through check for sending gift cards or money wire transfer.  \\ 
\textbf{Mortgage Scam} 
Mortgage Scams on the rise targeting desperate homeowners to be scammed  \cite{PleasantGreen2019}  Scammers may create a fake mortgage company website or claim to be calling from mortgage company to engage with homeowners for best offers, upfront costs, or deeds payment \cite{Union2020}. The scammer may collect the owner’s information from third-party or public search internet. For instance, scammer may tell the financially unstable homeowners to pay the upfront cost so that the company manages to pay the installments rest of the months until the homeowner able to pay the money back. If the homeowner pays the money through wire transfer, the scammer will disappear with the money and the homeowner victim will never be able to reach out to them for payments or get the money back. The scammers may shut down the fake company website domain or stop using the phone numbers used to contact the victims. If the scammer is an international caller, it is even much more complex to track the scammer and too late to respond to the scam.  \\ 
\textbf{Recruitment Scam}
Recruitment scams focus on the desperate jobseekers to let them believe that the jobs are available with great returns, packages in few days \cite{PleasantGreen2019a}. These scams are carried out through internet by hosting fake job posting websites operated by scammers. When the jobseeker browse these websites, they are being directed to scammer contact details page or personal information filling page. The user is also lured with lucrative job offers and instruct them to pay the placement fees. If the desperate jobseekers believe that these are legit requests, they may pay the fees with the hope that job offer is in their hands. But, when the amount transferred to scammers, they will not respond and usually too late to receive money back from the scammers. Some scammers may also use alternative strategies as per the jobseeker social and financial status to lure the money during the recruitment or job scam. It is always recommended to use legitimate and official websites to apply for jobs and never pay the money to receive the job offer \cite{recruit}.   \\ 
\textbf{Technical Support Scam}
Technical support scams mainly target the people, who has limited knowledge on computer technologies. Scammers contact the users claiming to be from Microsoft or dell companies and offer help to eradicate computer virus or fix the computer performance issues  \cite{TonyRomm2019}. It is inherent that the companies never call to any customers to fixing an issue. The customer first need to reach out to the company support if the device has warranty. If the user not able to identify them as scammers, they follow the instructions provided by scammers to install the remote monitoring solutions like team viewer for troubleshooting the issue or removing the fake malware. Once they access the victim machine using remote access applications, they may encrypt all the data and ask for ransom to retrieve their information back. In other cases, they may request the tech support fee for resolving the issue. If the victim send money to the scammer operated accounts, it is very unlikely to get back money layer. They simply disconnect the calls or machines once they receive the payment. To make the users believe that they are infected with virus, they instruct them to browse a fake website controlled by them \cite{Rauti2017}. When the victim browser the website, a pop-up is displayed to instruct them to contact the scammer customer support phone number.  \\ 
\textbf{Law Firm Scam}
Scammers take advantage of the attorney reputation to impersonate them and collect the money from people on behalf of the attorneys \cite{Blackford2020}. For instance, scammer spoof the attorney email and send flood of emails to the people in the attorney county with a link to GoFundMe for raising money to a good cause. If the normal people, who thinks that it is actually posted by their beloved and respected attorney in the county, they may donate the money. This money is actually received by the scammers. On the other hand, the scammers can also cheat the attorneys to wire transfer the money in exchange to the money paid through check. The scammers claim to settle their issues with an attorney and pay the money through check as an advance to work on the case. They also demand to pay the additional money on the check in wire transfer before cash out the money using check in the bank. If the attorney is not aware of these scams, they may transfer the money to the scammer account and they disappear from the attorney radar.  \\ 
\textbf{Cat Scam}
Some people became a victim of cat/dog scams in recent times. Normally, people have a tendency to adopt the cats whenever possible and can bear the costs. Scammers post the online ads and host some websites luring the people to buy or adopt the cute and adorable cats pictures. If the people are not paid lot of attention towards this scam, they would order the cat online for shipping \cite{Roche2021}. The scammers charge them the cost and shipping, and even will not send the cats within the time frame. The scammers may even delay and further blackmail the buyers to book animal abusing cases. They demand even more money from the people and earn more money from the legitimate people. When the scammers receive the money, they shut down the domains and may start similar campaigns with some other domain names to continue their scamming operations. Unfortunately, the victims can not take any action to receive their money, as the scammer operating bank accounts internationally, and don’t leave some clues to find them staying anywhere in the world. \\ 
\textbf{Landlord Scam/Rental Scam}
The scammers post fake house renting post in social media or paper ads to let the people read them and future renters reach out to the scammers for house inquiry \cite{Park2017}. Furthermore, the scammers pursue the renters to convince them for renting the home and demand them to pay advance payment as part of the house renting confirmation. If the renter performs wire transfer to the scammer, the claimed landlord scammers won’t respond once the payment received. Scammers can also act as a renter and may collect money from the landlords by scamming them. Scammers reach out to the landlords and offer the willingness to take the lease. If the landlord asks for advance payment, the scammer trick them to pay through check with amount more than asked and later ask the landlord to return the money with wire transfer. The landlord will become a victim of the scam and scammers will disappear from the victim radar. \\ 
\textbf{Craigslist Scam}
Craigslist is a common web marketplace for sellers and buyers to exchange goods and perform trading. Scammers use Craigslist as a platform to lure victim with great offers and get money from them. Typically, scammer post a good deal Ad on the product or any item for sale. If any buyer wanted to buy the product, the scammer would offer them for a good price, and ask them to transfer money prior to shipping the product. The scammers may tell realistic stories to make the buyers believe them and perform a wire transfer before even receiving the product. When the money is transferred to the scammer related account, the Ad will be disappeared from the site, and they may create similar ads in the Craigslist to continue their scam operations \cite{Youngsam2020}. So, it is highly recommended looking for the originality of the product in the Ad and identity verification of the seller prior to buying through online portals like Craigslist. \\ 
\textbf{Ewhoring Scam}
As the pandemic hits the world since last year, most of the people spend lot of time in their home. The Ewhoring scams are reportedly increased during the pandemic, as the people living alone in home may spend more time on internet. Ewhoring scam involves targeting the young adults and let them pose with private videos or pictures when scammers talking to them with fake online profiles\cite{Pastrana2019}. The scammers record the private videos or pictures, and demand the money for not disclosing these videos in public. Although the scam may seem to be difficult, it is highly possible to lure the money from the people and get away from the victim. Hence, the online or social media users should pay attention to the new friend requests and should pay more attention while talking to stranger in online social media.  \\



\begin{table*}[!h]
\centering
\caption{Recent social engineering scams technology comparison} \label{T:priorreview}
\begin{tabular}{|l|l|l|l|}
\hline
\textbf{Scam Type}   & \textbf{Techniques}  & \textbf{Device/Applications}  & \textbf{Technologies Involved}\\ \hline
Gift card Scam \cite{RonnieTokazowski2019}& Phone SE, Online SE  & Phone  & Google Ads, toll-free service, Fake domains, Fake social profile accounts   \\ \hline
Romance Scam \cite{Whitty2015} & Phone SE, Online SE & Phone/laptop & Social Network Account, Fake email \\ \hline
IRS Scam \cite{Bidgoli2017}  & Phone SE, Robo Calling & Phone & VoIP   call service, International Call forwarding service \\ \hline
COVID Unemployment \cite{Team2020}  & Loophole in public website & Web  & Fake   Emails, Fake Bank Accounts    \\ \hline
Check Scam \cite{Brandon2009} & Online SE & Laptop/Phone & Fake   Email, Fake Phone calls   \\ \hline
Mortgage Scam \cite{Union2020}  & Phone SE, Online SE & Laptop/Phone  & Fake   Email, Fake Phone calls, Fake websites     \\ \hline
Recruitment   Scam \cite{PleasantGreen2019a}  & Phone SE, Online SE & Laptop/Phone & Google   Ads, Fake Phone calls, fake website domains  \\ \hline
Technical Support Scam \cite{TonyRomm2019} & Online SE, Phone SE & Computer, Laptop & Remote   Access software, Internet  \\ \hline
Law  Firm Scams  \cite{Blackford2020} & Online  SE & Web, Email service & Email   spoofing, fake emails \\ \hline
Cat Scam  \cite{Roche2021} & Online SE & Web  & Google Ads, fake website domains      \\ \hline
Landlord Scam/Rental \cite{Park2017} & Online SE & Web & Google   Ads, fake website domains, fake website applications  \\ \hline
Craigslist Scam \cite{Youngsam2020}   & Online SE   & Web      & Craigslist   Website      \\ \hline
eWhoring Scam \cite{Pastrana2019} & Online SE    & Web    & Internet/online   chat  \\ \hline
\end{tabular}
\end{table*}

The Table \ref{T:priorreview} illustrates the comparison of the SE scams based on the scammer techniques, device or applications involved and technology involved to execute the scams. We can clearly see that all the scams either rely on phone or online SE techniques except COVID unemployment scam. Additionally, these scams involve device or applications, which may include the phone, computer, laptop, email services, or web applications. So, the user must be conscious about scams when receiving the phone calls or browsing the web applications. The phone based scams such as gift card, IRS, romance, mortgage and recruitment scams can be mitigated if the phone identity is verified by the carrier networks. On contrary, the eWhoring and Craigslist, landlord scams are actively seen in the web application based scam category. The technologies such as email, domain, Google Ads, social network accounts, websites are mainly involved to perform these scams. Overall, based on the Table  \ref{T:priorreview}, we can say that the scams can be mitigated if the user is aware of the phone calls, in which whom they are talking to and only browsing safe applications with little efforts to identify the scams. However, the understanding of how businesses work and the relationship between different business entities is the key to distinguish the scam or legit actions. 

\begin{table*}[!h]
\centering
\caption{Recent social engineering scam detection signs comparison }\label{T:priorreview1}
\resizebox{\textwidth}{!}{%
\begin{tabular}{|l|l|l|l|l|}
\hline
\textbf{Scam Type}   & \textbf{SE Principles} & \textbf{Scammer   Location}   & \textbf{Scam   Initiated by} & \textbf{Entities Involved}  \\ \hline
Gift  card Scam \cite{RonnieTokazowski2019} & Greedy  & Nigeria,   India \cite{GaryWarner2018}  & Victim & Public Companies, individuals, Companies offering Gift cards    \\ \hline                        
Romance   Scam \cite{Whitty2015} & Liking,   & Nigeria,   India  & Scammer   & Social   Network Companies, individuals, \\ \hline
IRS   Scam \cite{Bidgoli2017} & Scared & India  \cite{MansiChoksi2017}  & Scammer           & US   Government, Individuals \\ \hline
COVID   Unemployment \cite{Team2020}  & -     & Nigeria      & Scammer           & US   Government, Individuals    \\ \hline
Check   Scam \cite{Brandon2009}  & Greedy      & Nigeria  & Scammer & Check Companies, Individuals, Mediators \\ \hline
Mortgage   Scam \cite{Union2020} & Greedy,   Scared  & Nigeria   & Scammer           & Homeowners        \\ \hline
Recruitment   Scam  \cite{PleasantGreen2019a}  & Social   Proof   &   -   & Victim            & Job   Seekers          \\ \hline
Technical   Support Scam \cite{TonyRomm2019}        & Authority,   Commitment  & India, Nigeria & Scammer  & Individuals, Companies like   Microsoft                                                    \\ \hline
Law Firm Scams \cite{Blackford2020} & Social   Proof   & Nigeria  & Scammer   & Attorneys                                                                                  \\ \hline
Cat  Scam  \cite{Roche2021}  & Liking,  & African Countries  & Victim   & Pet Owners, Pet lovers    \\ \hline
Landlord   Scam/Rental  \cite{Park2017} & Greedy,  Commitment  &  African Countries  & Scammer,   Victim & Homeowners                                                                                 \\ \hline
Craigslist   Scam  \cite{Youngsam2020} & Greedy    & Nigeria  & Victim            & Craigslist, Individuals                                                                    \\ \hline
eWhoring   Scam  \cite{Pastrana2019} & Liking, Social proof, Commitment & Nigeria  & Scammer,   Victim & Individuals, Social network   Companies                                                    \\ \hline
\end{tabular}%
}
\end{table*}

The Table \ref{T:priorreview1} describes the categorization of the scams based on the SE principle used by scammer, scammer location, scam initiated by victim or scammer and the entities involved. Most of the scams performed based on the human greediness to let the user become victim by himself. The liking principle is also used in the three scams to lure the innocent people. The majority of scams are initiated by the scammers. So, the attention is required when receiving a phone call from unknown number or email from unknown address. The gift card scam, recruitment and Craigslist scams are initiated by victims, who trust the source of getting scammer email or phone number. Hence, the users should be aware of scams when referring online sources or phone number or email address. Another way of identifying the scammers is to recognize their accent. We have determined that the most of the scammers either located from Nigeria and India. So, identifying the accent of the caller and comparing with the typical nation's people accent can be helpful to detect the scams. Furthermore, there were multiple stakeholders involved in the scams and tracking the scammers is much more complex once scammed. So, the precautions like security awareness and education should be taken to not fall a victim of these scams. \\

\begin{table*}[!h]
\centering
\caption{Social engineering scam age and location based  comparison}\label{T:priorreview2}
\begin{tabular}{|l|l|l|l|l|}
\hline
\textbf{Scam}              & \textbf{Age} & \textbf{Location } & \textbf{Victim}  & \textbf{International/National} \\ \hline
Gift   card Scam   \cite{RonnieTokazowski2019}        & Any            & Anywhere                     & Individuals      & International                  \\ \hline
Romance   Scam \cite{Whitty2015}            & Above   40              & Anywhere                     & Individuals      & International                  \\ \hline
IRS   Scam  \cite{Bidgoli2017}               & Any         & Anywhere                     & US   Individuals & USA                      \\ \hline
COVID   Unemployment \cite{Team2020} & Any         & Anywhere                     & US   Individuals & USA                   \\ \hline
Check   Scam  \cite{Brandon2009}             & Above 18             & Rural  areas                & US   Individuals & USA                    \\ \hline
Mortgage   Scam  \cite{Union2020}          & Above 18                  & Anywhere                     & US   Individuals & USA                       \\ \hline
Recruitment   Scam \cite{PleasantGreen2019a}      & 18-50                     & Urban                        & US   Individuals & USA                   \\ \hline
Technical   Support \cite{TonyRomm2019}  & Any              & Anywhere                     & US   Individuals & USA                   \\ \hline
Law   Firm Scams \cite{Blackford2020}          & Above 20                   & Anywhere                     & US   Individuals & USA                    \\ \hline
Cat   Scam  \cite{Roche2021}               & Any                   & Anywhere                     & US Individuals   & Pet   friendly countries       \\ \hline
Landlord   Scam \cite{Park2017}           & Above 18                    & Anywhere                     & US   Individuals & USA                    \\ \hline
\end{tabular}
\end{table*}

The SE scams are further studied to perform age and location based comparison, as shown in Table \ref{T:priorreview2}. It is clear that most of the scams discussed in the article targeted US citizens. Few of the scams like Gift card scam and Romance scam are targeted to people internationally. Further, we determine that the scams can be targeted to victims located in either urban or rural areas as long as the victim has phone number and Internet. Most of the scams targeted to people of any ages. However, some of the scams are specially focused on few age groups people. For instance, Romance scams mostly target women, who are older than 40.  So, the age based classification can be helpful for better understanding the different scams.\\

\section{Gift Card Scam Case Study}
\label{sec:giftcard}

Scammers have been leveraging various ways to steal the money from the victims. Most well-known method was to use fake bank accounts and diverting the money to bank accounts across the nations prior to scammer receive the money. However, there are still possibilities to get caught, in particular, leaving the traces of bank accounts. So, scammers are looking advanced ways to hide their identity when being involved in scam operations. For example, scammers steal the gift card numbers from the victims, who brought from commercial stores such as Walmart, Target, Apple Store, Amazon. \\
This section discusses one of the gift card scams happened recently focusing on travel industry customers across the world population and particularly focusing on US population as a case study. It also uncovers various social engineering techniques and technologies used by scammers to deceive the customers. \\
\begin{figure}[!h]	
\centering
\includegraphics[width=8 cm,height=6.5cm]{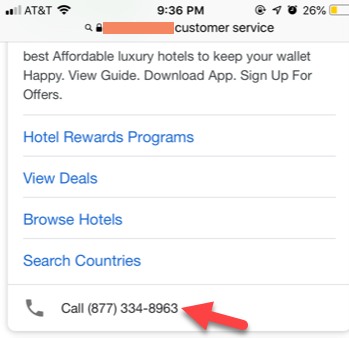}
\caption{Google Ads displaying fake customer contact phone number} 
\label{f.GoogleAds}
\end{figure}

Unlike popular scamming techniques like robocalling, phishing emails, Scammers leverage the public internet and popular web search engines like google search to post fake customer service numbers claiming to be travel company contact members. In general, the travel companies are usually not known about these fake customer service numbers distributing across the internet, as these are well crafted posts focused on the customer location as well as the time of the location with ephemeral in nature. In this case study, working for one of the leading travel companies in the world, we observed that the scammers were using the Google Ad services to register a fake travel consulting websites with maladvertising the scammer phone numbers as travel company customer service contact number. As shown in Figure \ref{f.GoogleAds}, when a user browses the internet and search for the travel company customer service number using Google Search, the first few results indexed in the Google search are related to the Google Ad Services and those ads hosts alike travel website with scammer fake phone number. When the user clicking the phone number (877) 334-8963 highlighted in the figure redirects the call to the scammer. \\
 
\begin{figure}[!h]	
\centering
\includegraphics[width=8 cm,height=7.5cm]{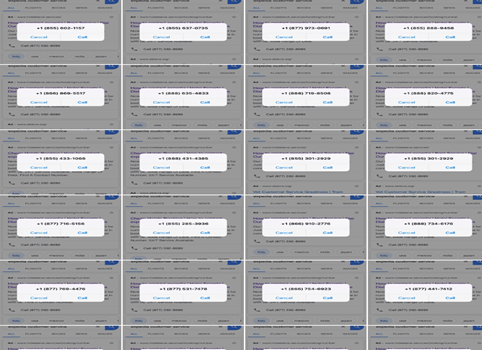}
\caption{Google Ads published phone number redirecting to multiple phone numbers}
\label{f.Phnumber}
\end{figure}

\begin{figure}[!h]	
\centering
\includegraphics[width=8 cm,height=6cm]{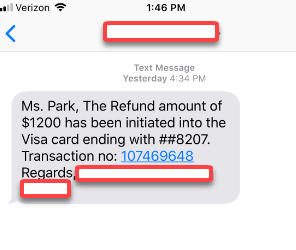}
\caption{Scammer sending phishing email to victim} 
\label{f.phishmessage}
\end{figure}

\begin{figure}[!h]	
\centering
\includegraphics[width=8cm,height=8cm]{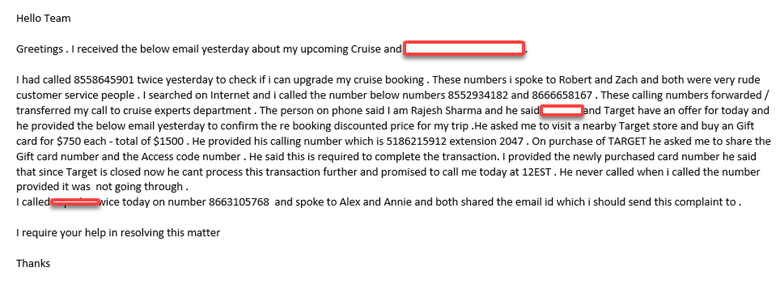}
\caption{Customer Email sample report } 
\label{f.customerreport}
\end{figure}

These fake phone numbers are appeared so random when the user search Google for customer service phone number and makes it even difficult to identify by Google teams as well as the customers trying to call the Expedia customer service with Google search phone number. As shown in the Figure \ref{f.Phnumber}, when click on the advertised phone number (877) 392-8999, it's redirecting to another toll-free number like (855) 802-1157 operated by scammers. When google search performed multiple times from a single machine and observed that more than 20 unique fake toll-free phone numbers appeared as seen in the Figure \ref{f.Phnumber}. \\
An innocent customer found the fake customer service phone number in internet would be contacted to scammers to inquire the travel itineraries including the travel cancellation, travel prepone or postpone, book a travel trip and other inquires. A scammer acts like travel company customer service representative and gather more details of the customer, possibly customer phone number, itinerary number and travel details using social engineering skills. Further, He/She makes fake claims saying that “currently great offer going on travel ticket sales and if you buy a gift card, you will have great discounts on travel booking” and lure the customers/victims to buy the gift card from public stores like Target, Walmart etc. They continue the conversation insist on sharing those bought gift card numbers and might use “greedy” tendency as a social engineering technique to pursue the gift card stealing during the call if victim is unwilling to provide the gift card. If the victim is disconnected during the call conversation, the scammer follows up with the victim through reverse voice call or message service, as shown in Figure \ref{f.phishmessage}. Once the gift card numbers shared with scammer, victims would be cutoff, and they are in dilemma of what happened to them. In this way, scammer use the technology and social engineering skills to pursue the scams in large scale across the nations and put the travel companies’ reputation in stake.\\
Few of the customers had started calling fake customer phone number and become a victim of spending money for gift cards. Later, we were seen the reports from customers claiming to fallen for these scams. A sample report is shown in Figure \ref{f.customerreport}. Based on our experience handling these gift card scams in real time, we provide the security incident and guidelines in the next section for effective mitigation of this scam.

\section{Incident Response}
\label{sec:incident}

In this section, we present the security incident response analysis to defend these scams in enterprise organizations following the NIST.SP.800-61r2 steps to handle an incident \cite{Cichonski2012}. The analysis contains how we have defended these scams in each step of the security incident life cycle as well as how the organizations get prepared to monitor and mitigate these social engineering scams. \\ 

\textbf{Preparation:}
The enterprise stakeholders such as marketing Ad management, Security incident and Response, Legal team, public relations team, Customer support team are trained to understand the roles and responsibility when the customer is being impacted with social engineering scams. Usually, the customer agents are being notify if the customer is impacted with the scam. The customer agent may engage or escalate to the security team for investigation. The Security team may work with other stakeholders as per necessity to take further actions. \\ 

\textbf{Detection and Analysis:}  
When the security incident response team is notified about the customer report. They may work and investigate the root cause of these reports. Most of the time, customer may not provide more information, in specific, how they are being impacted to lose the money. It may be important to work with the customers and get their feedback on the scam events happened. It also needed to collect the number of customer reports so that the impact and severity of the incident report can be determined, and actions taken accordingly. In most of the scenarios, the scammer collects the phone number and send a scam message to contact back to the scammer. An innocent customer may call the scammer assuming that they are talking to the company customer agent. So, the details like scammer phone number, voice recognition, behavior patterns, user location etc. can be obtained for tracking the scammers. If the root cause is found, for example, fake Google Ads in Gift card scheme, Security incident response team may interact with the team supports Google Ads to work with Google team for resolution. Google may assist to remove the fake Ads and implementing the future precautionary steps.  \\ 

\textbf{Containment, Eradication, and Recovery:}
The containment is difficult process in Gift card scams, as the Scammer already collected the Gift card number and It's too late to cancel gift cards. He/She may simply use those gift cards within few minutes, which leaves the customer or victim to have no idea on how it happened. However, the wide spread of the scams can be contained by taking precautionary measures and to fix the root cause. For instance, the fake google Ads created to reach out to the victim can be dismantled and may cancel the Google domain names. There is highly likely that the scammer come back with new google Ads. So, a constant pursues of taking down the Ads could permanently contain the gift card scams. We may also take additional countermeasures to monitor the Google Ads when type the relevant key work searches in google. Eradication involves further follow-ups with FTC and legal to report the evidence. They may take further actions to work with people in different country/state jurisdictions and coordinate to track down the scammer datacenters/locations to completely shut down their operations. But this requires much more efforts than we can think of. So, depends on the impact of the scam, the extent of the further actions is taken. Recovery step usually involves collecting the scammer stolen money from the customers. As the scammers operate from other countries in most of the time, it may be taken more than expected to recover the money if possible. If the notable number of people get impacted, then recovery efforts may be extensively needed. Overall, it depends on the scale of the scam, the recovery steps will be changes. \\ 

\textbf{Post-Incident Activity:}
Once the corresponding actions taken to mitigate the scam incident, The impacted customers are notified the root cause and let them sink in no involvement of the company. All the customers may be notified for identifying the scams and providing the awareness of how the scammer may leverage the company reputation for stealing the customer money. Further, a thorough investigation of the company misuse in social media and public internet and take the proper actions to remove those fake posts, advertisements to protect the people from not being a victim of these scams.

\section{Future Directions}
\label{sec:future}

\textbf{Security Awareness Training/Education}
The absolute eradication of the social engineering scams require educating the users on how to identify the different type of scams, providing the training on how to use the internet for web browsing, social network for communications and answering the phone calls without being impact by the robot or scammer calls. As the scammers mostly focused on the elder people and people living in rural areas, the scam awareness training should be provided the most probably impacting people. Furthermore, the phishing and scam identification education/training should be included in K-12 school curriculum to prepare the next generation for not being impacted with scams in the future.\\

\textbf{Data Analytics for scammer detection}
The scammer detection/tracking is a difficult task, as the little evidence leave by the scammers, and usually scammers operate from far from the victim nations. The jurisdiction rules and policies, Ownership issues on owning the scams for investigation when multiple stakeholders involved in the scams makes it even take a lot of time to resolve the scam cases. The Machine Learning/Artificial Intelligence (AI) technical solutions may need to be applied to detect the scam groups, behaviors in large-scale scams targeting the victims. The scammer advertisements or posts in the social network, scammer message conversations, scammer email conversations can be used as a dataset to apply data analytics algorithm and predict the scammer groups or locations and other information \cite{Vidros2017} \cite{Lansley2019} \cite{Lansley2020}. The applications of data analytics for scammer detection is one of the future directions going forward to mitigate the scams. \\

\textbf{Proactive Scam Information sharing}
The scammer move quickly to other scams and change their identities once the scam is  successful and the scammer is financially benefited. It is too late to identify the scammer, if the victim reports to the official authorities and started investigating the scams. The jurisdiction issue even makes it almost impossible to catch the scammers. It requires the mutual support from the two nations and should consider expediting the investigation process. So, technology may be used for proposing solutions to proactively sharing the information among the stakeholders and alerting the probably future targeting victims in advance. For instance, the blockchain technology may be used for secured data sharing among the stakeholders, and both private and public organizations react quickly to the scammer activity.\\

\textbf{Phone caller ID Authentication}
Most of the phone call based scams rely on robocalls to automate the tasks and let the victim respond to the scammer. The robocalls usually spoof the phone number and perform the calls to let the victim think that the calls are legitimate. The phone number identity authentication needs to be implemented to across all carriers to block the spoofed phone calls and mitigate some of the scam attempts. US FCC is working towards STIR/SHAKEN call ID authentication protocol implementation across the carriers and provided the guidelines for implementation \cite{Commission} FCC. However, there is no guarantee that the implementation may stop the spoofed phone calls, as the scammer may leverage the vulnerabilities in the STIR implementation in the carrier network and bypass the caller ID authentication. So, the proper implementation of the STIR protocol across the telecommunication industry is needed to prevent spoofed phone call based scams.

\section{Conclusion}
\label{sec:conclusion}

In this article, we have performed a review of the various social engineering scams including COVID-19 pandemic scams, which are mainly relying on the human persuasion by scammer rather than technological aspects like social engineering phishing email attacks. The existing works on the social engineering scam technical contributions are discussed and presented a social engineering attack/scam threat model architecture to represent any social engineering attack/scam with various core components or devices involved in the attack/scam life cycle. Additionally, two classic scams such as technical support scam and the COVID unemployment scam threat models are presented as an example. For scams awareness in organizations and individuals, we have leveraged the real time gift card scam targeting organization customers as a case study and presented a detail description of how the scam is executed by the scammers. Furthermore, the security incident response guidelines provided to prepare social engineering scams, which has nothing to do to the technical aspects and rather prepare the processes within the organization so that these scams can be prevented sooner and protect the organization reputation and customer base. We believe that our work can be used as a reference to continue technical research towards gaps in identifying the scams like cat scam, law firm, mortgage scam detection/prevention and proposing data analytics based scam detection solutions. Our work can also be used as a reference for anyone not to become a victim of the scams in the future.



{
\bibliographystyle{unsrt}
\bibliography{References}
}




\end{document}